\documentclass[aps,prl,twocolumn,superscriptaddress,showpacs]{revtex4}
\usepackage{graphicx}
\usepackage{dcolumn}
\usepackage{bm}
\usepackage{amsmath}

\begin{document}


\title{
$d$-$f$ Coulomb and quadrupole-strain interactions in DyB$_2$C$_2$ observed by resonant x-ray scattering
}


\author{T. Matsumura}
\email[]{tmatsu@iiyo.phys.tohoku.ac.jp}
\affiliation{Department of Physics, Graduate School of Science, Tohoku University, Sendai, 980-8578, Japan}
\author{D. Okuyama}
\affiliation{Department of Physics, Graduate School of Science, Tohoku University, Sendai, 980-8578, Japan}
\author{N. Oumi}
\affiliation{Department of Physics, Graduate School of Science, Tohoku University, Sendai, 980-8578, Japan}
\author{K. Hirota}
\affiliation{Institute for Solid State Physics, The University of Tokyo, Kashiwanoha, Kashiwa, 277-8581, Japan}
\author{H. Nakao}
\affiliation{Department of Physics, Graduate School of Science, Tohoku University, Sendai, 980-8578, Japan}
\author{Y. Murakami}
\affiliation{Department of Physics, Graduate School of Science, Tohoku University, Sendai, 980-8578, Japan}
\author{Y. Wakabayashi}
\affiliation{Institute of Materials Structure Science, High Energy Accelerator Research Organization, Tsukuba, 305-0801, Japan}


\date{\today}

\begin{abstract}
Experimental evidence of the $d$-$f$ Coulomb interaction responsible for the resonant x-ray scattering 
(RXS) from antiferroquadrupolar order in DyB$_2$C$_2$ is presented. The energy dependences of the RXS intensity 
with polarization analysis are analyzed by considering the interference between the resonances of dipolar ($E1$) 
and quadrupolar ($E2$) 
transition processes. It is found that the structure factors for the $E1$ and $E2$ processes have the same sign for 
$\sigma$-$\pi'$ but the opposite sign for $\sigma$-$\sigma'$ channel. This result, when compared with the calculated 
structure factors, means that the quadrupolar moments of the $4f$ and $5d$ electrons have opposite signs to each other. 
Interference between nonresonant Thomson scattering from the lattice distortion and the resonant scatterings is also 
studied and a direct coupling between $4f$ and the lattice is concluded. 
\end{abstract}


\maketitle


Resonant x-ray scattering (RXS) has been utilized extensively as a powerful tool to investigate periodic alignments of local 
anisotropies caused by, for example, chemical bondings and magnetic orderings~\cite{Templeton85, Gibbs88}. 
Element and electronic-shell selectivity provides valuable information on the anisotropy of a specific ion in solids. 
When the energy of an incident photon is tuned to an absorption edge of the element in study, 
a core electron is promoted into one of the unoccupied shells to form an intermediate state. 
When it decays back to the initial state, a secondary photon is emitted. 
The cross-section of this resonant process carries information on the anisotropy of the unoccupied shell. 
Since it was found in manganese oxides that RXS can also be a powerful tool to study orderings of 
electronic orbitals~\cite{Murakami98}, the method has also been applied to antiferroquadrupolar (AFQ) orderings in 
$f$ electron systems; e.g., in DyB$_2$C$_2$ and in CeB$_6$~\cite{Hirota00,Nakao01}.

Observations of the orbital orderings in Mn oxides by RXS have been performed to date mainly using the $E1$ resonance 
of the $K$ edge, which measures the anisotropy of the $4p$ state. Different theories have been proposed 
to explain the mechanism of the observation. 
One theory states that the $4p$ state reflects the anisotropy of the $3d$ state through Coulomb interaction~\cite{Ishihara98}. 
However, since the orbital orderings in transition-metal oxides generally accompany Jahn-Teller lattice 
distortions, the $4p$ states are affected also by the displacements of oxygen ions at the octahedral sites. 
Another theory states that this effect is much stronger than the Coulomb interaction 
because of the extended wavefunction of the $4p$ state~\cite{Elfimov99}. 
Since both theories give the same polarization and azimuthal-angle dependences, it has been difficult to distinguish 
the two interpretations. Recently, it was shown that the latter mechanism is dominant in Mn oxides from 
the experimental results on thin films~\cite{Ohsumi03}. 
In $4f$ electron systems, on the other hand, the relation among $4f$ orbital (quadrupolar moment), 
$5d$ orbital, and lattice has not yet been clarified. Three types of interactions should be considered: 
coupling between $4f$ and lattice, $5d$ and lattice, and $d$-$f$ Coulomb interaction. 
In the present paper we focus on the RXS from AFQ  order in DyB$_2$C$_2$.
A theoretical estimation of the RXS intensity claims that the effect of the lattice distortion of B and C atoms on the 
$5d$ state is much larger than the $d$-$f$ Coulomb interaction~\cite{Igarashi03}. 

\begin{figure}[tb]
\begin{center}
\includegraphics[width=7.5cm]{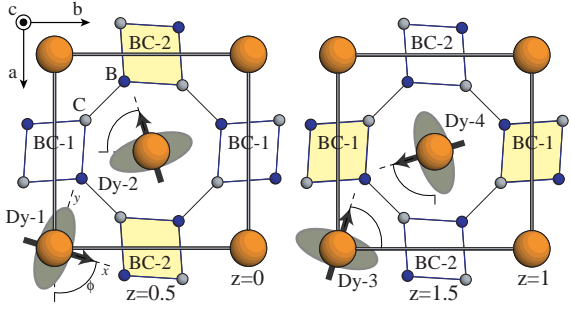}
\end{center}
\caption{Top view of the crystal structure of DyB$_2$C$_2$. 
Arrows indicate the magnetic moments in the AFM phase. Ellipses indicate the schematic views of 
pancake-like charge distributions of $4f$ electrons below $T_{Q}$. Their flat planes are perpendicular to the magnetic moments. 
The B-C layers are located at $z=0.5$. 
}
\label{fig1}
\end{figure}

In DyB$_2$C$_2$ the AFQ order takes place at $T_{Q}=24.7$ K, followed by an antiferromagnetic (AFM) order at 
$T_{N}=15.3$ K~\cite{Yamauchi99}. 
The magnetic structure below $T_{N}$ determined in Ref.~\onlinecite{Yamauchi99} is illustrated in Fig.~\ref{fig1}. 
The moments lie in the basal $c$ plane and those at the corner site and the face center site are coupled antiferromagnetically 
but canted from the $\langle$1 1 0$\rangle$ directions. The adjacent moments along the $c$ axis are almost 
perpendicular with each other. This structure is characterized by four $\bm{q}$-vectors: 
$\bm{q}_1$=(1 0 0), $\bm{q}_2$=(1 0 0.5), $\bm{q}_3$=(0 0 0), and $\bm{q}_4$=(0 0 0.5). 
This unusual magnetic structure can be understood only by considering the underlying AFQ order, which is 
illustrated by the ellipses in Fig.~\ref{fig1}. 

RXS experiment on DyB$_2$C$_2$ to observe the AFQ order 
was first performed by Hirota \textit{et al.} and by Tanaka \textit{et al.} using the 
$L_{\text{III}}$ edge of Dy~\cite{Hirota00}. 
Later, we made a detailed analysis on the observed resonant reflections using a theory developed 
by Blume~\cite{Matsumura02}. The calculated scattering amplitudes for each $\bm{q}$-vector, polarization channel 
($\sigma$-$\sigma'$ or $\sigma$-$\pi'$), 
and transition process ($E1$ or $E2$) assuming the AFQ and AFM structures 
shown in Fig.~\ref{fig1}, are mostly consistent with the experimental results. 
In short, the AFQ order is observed at $\bm{q}_4$ and $\bm{q}_2$, and a staggered lattice distortion at $\bm{q}_2$. 

However, in the previous works of RXS in DyB$_2$C$_2$, the (0 0 half-integer) reflections for the $\sigma$-$\sigma'$ 
channel in the $E2$ process are hidden in the tail of the $E1$ resonance, while those for the $\sigma$-$\pi'$ 
channel are well resolved. Since the calculated structure factors for the two channels have the same order of 
magnitudes, the disappearance of the $E2$ resonance in $\sigma$-$\sigma'$ has been a mystery. 
Recently, Tanaka \textit{et al.} made a systematic RXS study of the (0 0 half-integer) reflections and analyzed the data 
using the structure factor of the atomic tensors. They ascribed the disappearance to the idea that the rank 2 (quadrupole) 
and rank 4 (hexadecapole) tensors cancel out in the structure factor of the $\sigma$-$\sigma'$ channel~\cite{Tanaka04}. 

In the present paper, we report our new results of the RXS for the (0 0 2.5)$ (\in\bm{q}_4$) and (3 0 1.5) ($\in\bm{q}_2$) reflections 
and show that the above problem can be solved by considering the interference between the $E1$ and $E2$ resonances; 
the  $4f$-$5d$ Coulomb interaction will be revealed. Further, in the (3 0 1.5) reflection, 
the interference of the resonances with the Thomson scattering from the lattice distortion allows determination of 
the relationship between the orbitals and the lattice distortion; it will be shown that the lattice is coupled with the $4f$ orbital, and not with the 
$5d$ orbital. 

RXS experiments were performed at the beam line 16A2 of the Photon Factory, KEK. 
The incident energy was tuned to the $L_{\text{III}}$ absorption edge of Dy and the polarization of the 
diffracted beam was analyzed using the PG (0 0 6) reflection. Flat (0 0 1) and (2 0 1) surfaces of the DyB$_2$C$_2$ samples 
were prepared for the two experiments on the (0 0 2.5) and (3 0 1.5) reflections, respectively. 
The azimuthal angle $\Psi$ is defined to be zero when the $b$ axis is along $\bm{k}+\bm{k}'$, the vector sum of the 
incident and scattered x rays.

\begin{figure}[tb]
\begin{center}
\includegraphics[width=7.8cm]{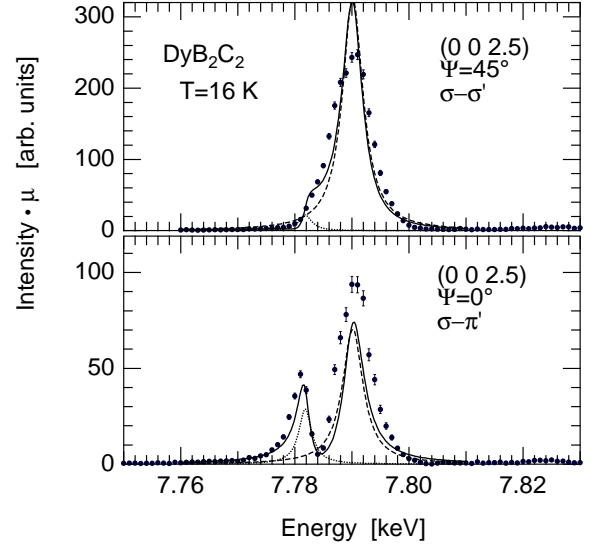}
\end{center}
\caption{Incident energy dependences of the intensity of the (0 0 2.5) reflection in the AFQ phase for the $\sigma$-$\sigma'$
channel at $\Psi=45^{\circ}$ and $\sigma$-$\pi'$ channel at $\Psi=0^{\circ}$. The data are corrected for the absorption. 
Solid lines are the fits considering the interference between $E1$ and $E2$ resonances. 
Dotted and dashed lines are the contributions from $E1$ and $E2$ processes, respectively, without the interference. 
}
\label{fig2}
\end{figure}

Figure~\ref{fig2} shows the energy dependences of the intensity of the (0 0 2.5) reflection in the AFQ phase for 
$\sigma$-$\sigma'$ and $\sigma$-$\pi'$ channels. Solid lines are the fits with the structure factor
\begin{equation}
F_{\lambda\lambda'}=\frac{F^{(E2)}_{\lambda\lambda'}}{E-\Delta_{E2}+i\Gamma_{E2}/2}
+r\frac{ F^{(E1)}_{\lambda\lambda'}}{E-\Delta_{E1}+i\Gamma_{E1}/2}\; ,
\label{eq:1}
\end{equation}
where $|F_{\lambda\lambda'}|^2$ were compared with the data. 
The parameters obtained are the followings: $F^{(E2)}_{\sigma\sigma'}=-0.00647$, $F^{(E1)}_{\sigma\sigma'}=0.0396$, 
$F^{(E2)}_{\sigma\pi'}=0.00738$, $F^{(E1)}_{\sigma\pi'}=0.0184$, $\Delta_{E2}=7.782$ keV, 
$\Delta_{E1}=7.790$ keV, $\Gamma_{E2}=2.75$ eV, and $\Gamma_{E1}=4.39$ eV. The mixing parameter $r$, 
introduced in Ref.~\onlinecite{Tanaka04}, was fixed at unity, which means that the $E1$ and $E2$ resonances interfere coherently. 
It should be noted that the data can be explained with finite $F^{(E2)}_{\sigma\sigma'}$ 
of the same order of magnitude as $F^{(E2)}_{\sigma\pi'}$ if the sign of $F^{(E2)}_{\sigma\sigma'}$ 
is opposite to that of $F^{(E1)}_{\sigma\sigma'}$; this is due to the interference effect.

The calculated structure factors of the even rank atomic tensor for the (0 0 half-integer) reflections, using the formalism 
in Refs.~\onlinecite{Lovesey96} and ~\onlinecite{Lovesey01}, are written as 
\begin{eqnarray}
F^{(E1)}_{\sigma\sigma'}&=& A \sin 2\Psi \sin 2\phi \langle T^{(2)}_{2}\rangle_{5d} \;, \\
F^{(E1)}_{\sigma\pi'}&=& A \cos 2\Psi \sin\theta 
       \sin 2\phi \langle T^{(2)}_{2}\rangle_{5d} \;,
\end{eqnarray}
for the $E1$ process and 
\begin{eqnarray}
F^{(E2)}_{\sigma\sigma'}&=& B \sin2\Psi \sin^2\theta \sin 2\phi \nonumber \\
  && \times \{3\sqrt{2}\langle T^{(2)}_{2}\rangle_{4f} - \sqrt{11} \langle T^{(4)}_{2}\rangle_{4f} \} \;, \\
F^{(E2)}_{\sigma\pi'}&=& -B \cos2\Psi \sin\theta \sin 2\phi 
   \{ 3\sqrt{2} \langle T^{(2)}_{2}\rangle_{4f} (3-4\sin^2\theta)  \nonumber \\
        && + \frac{\sqrt{11}}{2} \langle T^{(4)}_{2}\rangle_{4f}  (1+\sin^2\theta)\} \;, 
\end{eqnarray}
for the $E2$ process, where $\sin^2\theta=0.315$ for (0 0 2.5). $A$ and $B$ are the positive constant factors 
in $F^{(E1)}$ and $F^{(E2)}$, respectively. Note that the $E1$ and $E2$ resonances have the same azimuthal-angle 
dependence for both polarization channels. 
$\langle T^{(K)}_q \rangle$ represents the matrix element of the $q$th component 
of the atomic tensor of rank $K$ in the local ionic coordinates, where the principal axis of the quadrupolar moment
is taken as the $x$ axis and the $c$ axis of the crystal as the $z$ axis. 
The $x$ axis coincides with the direction of the magnetic moment in the AFM phase, 
which is canted from the $\langle$1 0 0$\rangle$ directions by an angle $\phi$. 
Since the ion is in the local symmetry of $2/m$ in the AFQ phase~\cite{Tanaka04,Adachi02}, 
$\langle T^{(K)}_q \rangle$ with $K$=even has only the components $q=\pm 2$ 
and the relation $\langle T^{(K)}_2 \rangle=\langle T^{(K)}_{-2} \rangle$ holds. Both $\langle T^{(2)}_2 \rangle$ and 
$\langle T^{(4)}_2 \rangle$ are real and $\langle T^{(2)}_2 \rangle$ represents the $(x^2-y^2)$-type 
quadrupole moment and $\langle T^{(4)}_2 \rangle$ represents the $\{x^4-y^4-\frac{6}{7}(x^2-y^2)r^2\}$-type 
hexadecapole moment. It should be noted that $\langle T^{(K)}_q \rangle$ in the structure factors of 
$E1$ and $E2$ processes represent atomic tensors of the $5d$ and $4f$ shells, respectively, 
which is indicated by the suffix.

If the sign of $\langle T^{(2)}_ 2\rangle_{5d}$ is opposite to that of $\langle T^{(2)}_ 2\rangle_{4f}$, 
we have opposite signs for $F^{(E2)}_{\sigma\sigma'}$ and $F^{(E1)}_{\sigma\sigma'}$ and the 
same signs for $F^{(E2)}_{\sigma\pi'}$ and $F^{(E1)}_{\sigma\pi'}$.  $F^{(E2)}_{\sigma\sigma'}$ and 
$F^{(E2)}_{\sigma\pi'}$ have the same order of magnitudes if we ignore the rank 4 tensors. 
This is consistent with the fitting result of the data and can explain the obscure and distinct 
features of the $E2$ resonance in the $\sigma$-$\sigma'$ and $\sigma$-$\pi'$ channels, respectively. 
The present analysis can also explain the experimental results in Ref.~\onlinecite{Tanaka04}. 
Indeed, the data for $\sigma$-$\sigma'$ can actually be fitted with vanishing $E2$ term as claimed in 
Ref.~\onlinecite{Tanaka04} if we use larger $\Gamma_{E1}$ for $\sigma$-$\sigma'$. However, 
the rank 4 term in $F^{(E2)}$ is not necessary to explain the result. This is justified by the azimuthal-angle 
dependence of the $E2$ resonance for the (3 0 1.5) reflection, where only the feature of the rank 2 tensor is observed with  
no indication of the rank 4 tensor; this point will be presented in detail in a forthcoming paper.

The above analysis also applies to the RXS of the (3 0 1.5) reflection, where a large nonresonant Thomson scattering from the lattice distortion also interferes. 
Concerning the lattice distortion below $T_Q$, Adachi \textit{et al.} recently presented firm evidence that B and C atoms are shifted along the 
$c$ axis~\cite{Adachi02}. Here, two cases can be considered. In the first case the BC-1 parallelogram in Fig.~\ref{fig1} is shifted up (down) 
and the BC-2 is shifted down (up) at $z=$0.5 (1.5), and in the second case vice versa. 
The structure factor $F_{\text{lat}}$ for the two cases are the same in 
magnitude but opposite in sign. Although the sign cannot be determined by normal Thomson scattering, 
analysis of the interference with the resonance makes it possible as described next. 

\begin{figure}[tb]
\begin{center}
\includegraphics[width=7.8cm]{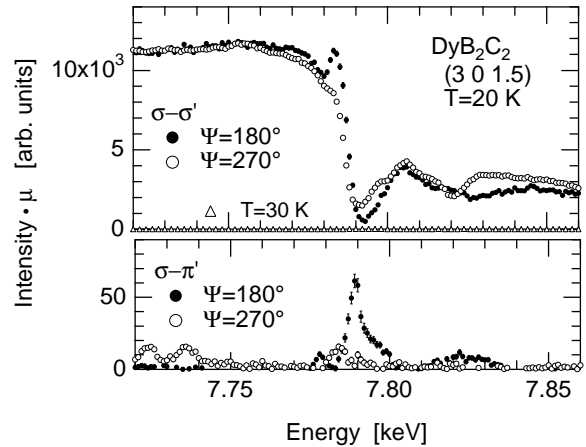}
\end{center}
\caption{Incident energy dependences of the intensity of the (3 0 1.5) reflection in the AFQ phase at $\Psi=180^{\circ}$ 
and $\Psi=270^{\circ}$. The data are corrected for the absorption, although it is not perfect. The intensities well 
below and well above the edge should be equal if the correction is perfect. 
}
\label{fig3}
\end{figure}

Figure~\ref{fig3} shows the incident energy dependence of the intensity of the (3 0 1.5) reflection. 
The nonresonant region far below the absorption edge do not exhibit azimuthal-angle dependence, while 
the intensities at energies of $E1$ and $E2$ resonances vary with the azimuthal angle, indicating that there 
are indeed resonances. 

\begin{figure}[tb]
\begin{center}
\includegraphics[width=7.8cm]{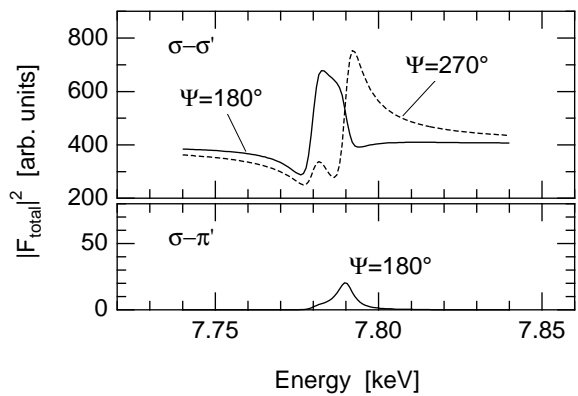}
\end{center}
\caption{Simulation of the energy dependence of the intensity of the (3 0 1.5) reflection at $\Psi=180^{\circ}$ and 
$\Psi=270^{\circ}$. The parameters are taken to be $F_{\text{lat}}=20$, $F^{(E2)}=F^{(E2)}_{\text{calc}}\times (-0.1)$,  
$F^{(E1)}=F^{(E1)}_{\text{calc}}\times 0.033$, $\Gamma_{E1}=\Gamma_{E2}=6$ eV, 
$\Delta_{E2}=7.78$ keV, and $\Delta_{E1}=7.79$ keV. The rank 2 structure factors for the $\sigma$-$\pi'$ channel 
vanishes at $\Psi=270^{\circ}$. 
}
\label{fig4}
\end{figure}

We have calculated the structure factors of the rank 2 atomic tensor for the (3 0 1.5) reflection and have 
found that the sign of $\langle T^{(2)}_{2}\rangle_{5d}$ must be opposite to that of $\langle T^{(2)}_{2}\rangle_{4f}$ 
to explain the result in Fig.~\ref{fig3}, which is consistent with the (0 0 2.5) reflection. Further, in order to explain the 
resonances of $E2$ and $E1$ processes in the $\sigma$-$\sigma'$ channel, where the Thomson scattering 
also interferes, the sign of $F_{\text{lat}}$ must be opposite to that of $\langle T^{(2)}_{2}\rangle_{4f}$ 
because the coefficient of $\langle T^{(2)}_{2}\rangle_{4f}$ in $F^{(E2)}_{\sigma\sigma'}$ is negative. 
This means that the first case of the shift of the B-C parallelograms is realized, and not the second case. 
A simulation of the interference among nonresonant Thomson scattering, $E2$ resonance, and $E1$ resonance  
is demonstrated in Fig.~\ref{fig4}. $F_{\text{calc}}$ represents the coefficient of $\langle T^{(2)}_{2}\rangle$ in the 
calculated structure factors. 

It must be noticed that the matrix element of $\langle T^{(2)}_2 \rangle$ involves the reduced matrix element, 
which contains information of the intermediate state and can be positive or negative depending on the ion species. 
When the charge distribution of $4f$ electrons of a Dy ion is extended along the $y$ direction of the local coordinates, 
the quadrupolar moment $O_{22}$ $(=J_x^{\;2}-J_y^{\;2})$ is positive. 
The matrix element $\langle T^{(2)}_2 \rangle_{4f}$ then 
becomes negative because the reduced matrix element of Dy$^{3+}$ for rank 2 is negative~\cite{Lovesey96}. Non-zero 
$\langle T^{(2)}_2 \rangle_{5d}$ value is caused by the lifting of the degeneracy of $yz$ and $zx$ type 
$5d$ orbitals in the tetragonal environment. 
The local lattice distortion of B and C atoms in the present case will favor the $yz$ type orbital, 
which extends along the $y$ direction and has less mixing with the $2p$ orbitals of B and C than 
the $zx$ type orbital. However, the calculation of $\langle T^{(2)}_2 \rangle_{5d}$ for the $yz$ 
orbital, following the formalism described in Ref.~\onlinecite{Lovesey96}, gives a negative value, 
whereas $\langle T^{(2)}_2 \rangle_{5d}$ becomes positive for the 
$zx$ orbital. Then, the result of the analysis that the signs of $\langle T^{(2)}_2 \rangle_{4f}$ and 
$\langle T^{(2)}_2 \rangle_{5d}$ are opposite indicates that the $zx$ orbital is favored in the $5d$ state. 

The present analysis assumes localized atomic states both for $4f$ and $5d$ states with fixed energy levels 
in the framework of the \textit{idealized scattering length} ~\cite{Lovesey96}. 
We have also assumed that $F_{\text{lat}}$, $F^{(E2)}$, and $F^{(E1)}$ interfere coherently. 
Actually, a resonance is composed of many intermediate states with slightly different energies. 
Although the consistent explanation of the data supports these assumptions, it should be 
examined by more realistic theoretical calculations that take into account complex energy levels of the intermediate 
state~\cite{Igarashi03}. 

In conclusion, 
the charge distributions of the $4f$ and $5d$ electrons of Dy are extended along the directions that are orthogonal with each other 
because of the $d$-$f$ Coulomb interaction. This result is consistent with the recent re-examination of the RXS 
intensity~\cite{Nagao03}. 
The relation with the lattice distortion is such that the B-C atoms in the direction where the $4f$ charge distribution 
is extended move away from Dy, while those in the direction where the $5d$ charge distribution is extended 
move close to Dy. 
Although the $5d$ orbital seems to be strongly coupled with the lattice, the present result indicates that the $4f$ orbital is 
much more strongly coupled with the lattice. The resultant lattice distortion is such that the quadrupole-strain interaction 
of the $4f$ electrons gains energy, overwhelming the coupling between $5d$ and the lattice. 
This may be because the $yz$ and $zx$ orbitals in the $5d$ state are almost empty and they have little effect 
on the total energy when the lattice is distorted. Energy band calculation of LaB$_2$C$_2$ shows that the 
conduction band of the La $5d$ state consists of the $xy$ orbital~\cite{Harima}. It is expected that 
a small amount of $zx$ component is induced in the AFQ phase by the $d$-$f$ Coulomb interaction, giving rise to 
the $E1$ resonance.

The authors acknowledge Prof. S. W. Lovesey for valuable discussions and comments on the application of his theory 
to the analysis of our experimental data. We also thank Prof. H. Harima for detailed information on the result of 
band calculation.  This study was performed under the approval of the Photon Factory Program Advisory 
Committee (Proposal No. 2001G063), and was supported by a Grant-in-Aid for Scientific Research from the 
Japanese Society for the Promotion of Science.


\end{document}